# Time-Dependent Quantum Transport Through Graphene Nanoribbons

Hang Xie[1], Yanho Kwok[1], Yu Zhang[1], Feng Jiang[2], Xiao Zheng[3], YiJing Yan[2,3] and GuanHua Chen[1,*]

[1] *Department of Chemistry, The University of Hong Kong, Hong Kong*

[2] *Department of Chemistry, Hong Kong University of Science and Technology, Hong Kong*

[3] *Hefei National Laboratory for Physical Sciences at the Microscale, University of Science and Technology of China, Hefei 230026, China*

## Abstract

Time-dependent quantum transport for graphene nanoribbons (GNR) are calculated by the hierarchical equation of motion (HEOM) method based on the nonequilibrium Green's function (NEGF) theory (Xie et.al, J. Chem. Phys. 137, 044113, 2012). In this paper, a new steady state calculation technique is introduced and accelerated by the contour integration, which is suitable for large systems. Three Lorentzian fitting schemes for the self-energy matrices are developed based on the nonlinear least square method. Within these schemes, the number of Lorentzians is effectively reduced and the fitting results are good and convergent. With these two developments in HEOM, we have calculated the transient currents in GNR. We find a new type of edge state with delta-function-like density of states in many semi-infinite armchair-type GNR.

## I Introduction

Quantum transport is an important field of research, due to the rapid improvement in nanotechnology and the semiconductor industry [1-3]. First-principles nonequilibrium Green's function (NEGF) theory has been widely employed to simulate the steady state currents through the electronic devices [4-6]. Besides the static calculations, the time-dependent quantum transport theory has been developed within the framework of NEGF[7-8] and time-dependent density functional theory (TDDFT)[9-13]. In time-dependent quantum transport calculations, the wide band limit (WBL) approximation is often used [7,13]. Recently we have developed a first-principle HEOM for the reduced single-electron density matrix (RSDM) (denoted as RSDM-HEOM) [14-18]. Multi-Lorentzian expansion is employed to approximate the lead spectra matrices[17]. As RSDM is much simpler than many-electron density matrix, the RSDM-HEOM method is much more efficient so that it is employed to model the realistic systems. In the HEOM calculation for large systems, it is difficult to calculate the initial values[17]. And the computational load of HEOM is related to the number of Lorentzians. For large systems, it is crucial to find a good set of Lorentzians, which mimic the lead spectra profiles with minimal number of Lorentzians.

In this paper, we derive a new integral formula for the initial values. The initial HEOM values of very large systems can be easily calculated. A set of Lorentzian fitting algorithms is developed. The least square method is used in the fitting. These Lorentzian fitting schemes automatically fit all the linewidth function curves with a small number of Lorentzians. This manuscript is organized as follows. The methodology is described in Sec. II, including the HEOM, the initial value calculations, and three new schemes of Lorentzian fitting algorithms. In Sec. III, numerical results for the graphene nanoribbons are given. Two types of nanoribbons (armchair and zigzag) are investigated. Summary is given in Sec. IV.

## II Methodology

The HEOM method is based on the non-equilibrium Green's function theory. Instead of calculating the time evolution of the Green's functions, this method defines some time integrations of the Green's functions and self-energies as the auxiliary density matrices (ADM). Then with the equation of motions of the Green's functions and self-energies, a set of time differential equations for the density matrices and auxiliary density matrices are derived. Sec. II A gives the details for this method.

### A. Introduction to HEOM

In the lead-device-lead system, the equation of motion for the device's density matrix is given below [13],

$$i\dot{\boldsymbol{\sigma}}_D(t) = [\mathbf{h}_D(t), \boldsymbol{\sigma}_D(t)] + i\int_{-\infty}^{t} d\tau [\mathbf{G}_D^<(t,\tau) \cdot \boldsymbol{\Sigma}_\alpha^>(\tau,t) - \mathbf{G}_D^>(t,\tau) \cdot \boldsymbol{\Sigma}_\alpha^<(\tau,t) - H.C.] \quad (1)$$

where $\boldsymbol{\sigma}_D$ and $\mathbf{h}_D$ are the density matrix and Hamiltonian of device respectively. $\boldsymbol{\Sigma}_\alpha^x(t,\tau)$ are the lesser (x=<) or greater (x=>) self-energy; $\mathbf{G}_D^x(t,\tau)$ are the lesser or greater Green's function of device. *H.C.* means the Hermitian conjugate;

According to reference [16], if we introduce the energy-resolved self-energies $\boldsymbol{\Sigma}_\alpha^{<,>}(\varepsilon,\tau,t)$, ($\boldsymbol{\Sigma}_\alpha^{<,>}(\tau,t) = \int d\varepsilon \cdot \boldsymbol{\Sigma}_\alpha^{<,>}(\varepsilon,\tau,t)$), and the following auxiliary density matrices (AMD) of the 1$^{st}$ and 2$^{nd}$ tier,

$$\boldsymbol{\varphi}_\alpha(\varepsilon,t) = i\int_{-\infty}^{t} d\tau [\mathbf{G}_D^<(t,\tau) \cdot \boldsymbol{\Sigma}_\alpha^>(\varepsilon,\tau,t) - \mathbf{G}_D^>(t,\tau) \cdot \boldsymbol{\Sigma}_\alpha^<(\varepsilon,\tau,t)] \quad , \quad (2)$$

$$\boldsymbol{\varphi}_{\alpha\alpha'}(\varepsilon,\varepsilon',t) = i\int_{-\infty}^{t} dt_1 \int_{-\infty}^{t} dt_2 \{[\boldsymbol{\Sigma}_{\alpha'}^<(\varepsilon',t,t_1) \cdot \mathbf{G}_D^a(t_1,t_2) + \boldsymbol{\Sigma}_{\alpha'}^r(\varepsilon',t,t_1) \cdot \mathbf{G}_D^<(t_1,t_2)]\boldsymbol{\Sigma}_\alpha^>(\varepsilon,t_2,t)$$
$$-[\boldsymbol{\Sigma}_{\alpha'}^>(\varepsilon',t,t_1) \cdot \mathbf{G}_D^a(t_1,t_2) + \boldsymbol{\Sigma}_{\alpha'}^r(\varepsilon',t,t_1) \cdot \mathbf{G}_D^>(t_1,t_2)]\boldsymbol{\Sigma}_\alpha^<(\varepsilon,t_2,t)\} , \quad (3)$$

we may derive the following set of equations,

$$i\dot{\boldsymbol{\sigma}}_D(t) = [\mathbf{h}_D(t), \boldsymbol{\sigma}_D(t)] - \sum_\alpha \int d\varepsilon \cdot [\boldsymbol{\varphi}_\alpha(\varepsilon,t) - \boldsymbol{\varphi}^\dagger_\alpha(\varepsilon,t)] . \quad (4)$$

$$i\dot{\boldsymbol{\varphi}}_\alpha(\varepsilon,t) = [\mathbf{h}_D(t) - \varepsilon - \Delta_\alpha(t)] \cdot \boldsymbol{\varphi}_\alpha(\varepsilon,t) + [f_\alpha(\varepsilon) - \boldsymbol{\sigma}_D(t)]\boldsymbol{\Lambda}_\alpha(\varepsilon) + \sum_{\alpha'}^{N_\alpha} \int d\varepsilon' \boldsymbol{\varphi}_{\alpha\alpha'}(\varepsilon,\varepsilon',t) \quad , \quad (5)$$

$$i\dot{\boldsymbol{\varphi}}_{\alpha,\alpha'}(\varepsilon,\varepsilon',t) = -[\varepsilon + \Delta_\alpha(t) - \varepsilon' - \Delta_{\alpha'}(t)] \cdot \boldsymbol{\varphi}_{\alpha,\alpha'}(\varepsilon,\varepsilon',t)$$

$$+\mathbf{\Lambda}_{\alpha'}(\varepsilon')\cdot\mathbf{\varphi}_\alpha(\varepsilon,t)-\mathbf{\varphi}^\dagger_{\alpha'}(\varepsilon',t)\cdot\mathbf{\Lambda}_\alpha(\varepsilon) \quad . \tag{6}$$

where $\mathbf{\Lambda}_\alpha(\varepsilon)$ is the linewidth function, and $f_\alpha(\varepsilon)$ is the Fermi function for lead $\alpha$, $f_\alpha(\varepsilon)=1/(1+\exp[\beta(\varepsilon-\mu_\alpha)])$, $\beta=1/k_BT$, is the reciprocal temperature, , and $\mu_\alpha$ is the chemical potential of lead $\alpha$. $\Delta_\alpha(t)$ is the time-dependent bias potential in lead $\alpha$. This is the RSDM-HEOM. The derivation details of Eqs. (4)-(6) are given in reference [16-17].

In Eqs.(4) and (5), there are energy integrations for $\mathbf{\varphi}_\alpha(\varepsilon,t)$ and $\mathbf{\varphi}_{\alpha\alpha'}(\varepsilon,\varepsilon',t)$, which is very computationally expansive. In numerical calculations, the residue theory is used to transfer these integrals into summations [17]. Here we show the brief steps.

Step (1), the steady state self-energy $\mathbf{\Sigma}^{<,>}_\alpha(\tau,t)$ can be transferred into summation by the residue theory. For example,

$$\mathbf{\Sigma}^<_\alpha(\tau-t)=\frac{i}{2\pi}\int_{-\infty}^{+\infty}f_\alpha(\varepsilon)\mathbf{\Lambda}_\alpha(\varepsilon)\cdot e^{-i\varepsilon(\tau-t)}d\varepsilon=\frac{i}{2\pi}\oint f_\alpha(z)\mathbf{\Lambda}_\alpha(z)\cdot e^{-iz(\tau-t)}dz. \tag{7}$$

With the multi-Lorentzian expansion to estimate the linewidth function ($\mathbf{\Lambda}_\alpha(\varepsilon)\approx\sum_{d=1}^{N_d}\frac{\eta_d}{(\varepsilon-\Omega_d)^2+W_d^2}\overline{\mathbf{\Lambda}}_{\alpha d}$), and the Padé expansion to approximate the Fermi-Dirac function [19] ($f_\alpha(z)=\frac{1}{1+\exp(z)}\approx\frac{1}{2}+\sum_{p=1}^{N_p}(\frac{R_p}{z-z_p^+}+\frac{R_p}{z-z_p^-})$), we may find the poles of the integrand in the integration contours and write the residue summation form of this integral:

$$\mathbf{\Sigma}^{<,>}_\alpha(\tau-t)=\sum_k^{N_k}\mathbf{A}^{<,>\pm}_{\alpha k}\cdot e^{\mp\gamma^-_{\alpha k}(t-\tau)} \quad .$$

The expressions for $\mathbf{A}^{<,>\pm}_{\alpha k}$ and $\gamma^\mp_{\alpha k}$ are given in reference [17]. '+' and '−' correspond to different contours.

Step (2), adding a phase factor to Eq. (7), $\mathbf{\Sigma}^{<,>}_\alpha(\tau,t)$ can also be written as a summation form:

$$\mathbf{\Sigma}^{<,>}_\alpha(t-\tau)=e^{-i\int_\tau^t\Delta_\alpha(\xi)d\xi}\cdot\mathbf{\Sigma}^{<,>}_\alpha(\tau-t)=\sum_{k=1}^{N_k}\mathbf{\Sigma}^{<,>}_{\alpha k}(t-\tau). \tag{8}$$

Then the integral of $\mathbf{\varphi}_\alpha(\varepsilon,t)$ can be transferred into summation,

$$\int d\varepsilon\cdot\mathbf{\varphi}_\alpha(\varepsilon,t)=\int d\varepsilon\cdot i\int_{-\infty}^t d\tau[\mathbf{G}^<_D(t,\tau)\cdot\mathbf{\Sigma}^>_\alpha(\varepsilon,\tau,t)-\mathbf{G}^>_D(t,\tau)\cdot\mathbf{\Sigma}^<_\alpha(\varepsilon,\tau,t)]=\sum_{k=1}^{N_k}\mathbf{\varphi}_{\alpha k}(t) \quad , \tag{9a}$$

where

$$\boldsymbol{\varphi}_{\alpha k}(t) = i \int_{-\infty}^{t} d\tau [\mathbf{G}_D^<(t,\tau) \cdot \boldsymbol{\Sigma}_{\alpha k}^>(\tau,t) - \mathbf{G}_D^>(t,\tau) \cdot \boldsymbol{\Sigma}_{\alpha k}^<(\tau,t)] \tag{9b}$$

is called the discretized 1$^{st}$ tier auxiliary density matrix.

Step (3), similarly, the integral of $\boldsymbol{\varphi}_{\alpha\alpha'}(\varepsilon,\varepsilon',t)$ can be expressed into summation

$$\iint d\varepsilon d\varepsilon' \cdot \boldsymbol{\varphi}_{\alpha\alpha'}(\varepsilon,\varepsilon',t) = \sum_{k,k'}^{N_k} \boldsymbol{\varphi}_{\alpha k,\alpha'k'}(t) \tag{10a}$$

with each discrete term defined as

$$\boldsymbol{\varphi}_{\alpha k,\ \alpha'k'}(t) = i \int_{-\infty}^{t} dt_1 \int_{-\infty}^{t} dt_2 \{[\boldsymbol{\Sigma}_{\alpha'k'}^<(t,t_1) \cdot \mathbf{G}_D^a(t_1,t_2) + \boldsymbol{\Sigma}_{\alpha'k'}^r(t,t_1) \cdot \mathbf{G}_D^<(t_1,t_2)]\boldsymbol{\Sigma}_{\alpha k}^>(t_2,t)$$
$$- [\boldsymbol{\Sigma}_{\alpha'k'}^>(t,t_1) \cdot \mathbf{G}_D^a(t_1,t_2) + \boldsymbol{\Sigma}_{\alpha'k'}^r(t,t_1) \cdot \mathbf{G}_D^>(t_1,t_2)]\boldsymbol{\Sigma}_{\alpha k}^<(t_2,t)\}. \tag{10b}$$

From these definitions, the following central equations can be derived

$$i\dot{\boldsymbol{\sigma}}(t) = [\mathbf{h}_D(t), \boldsymbol{\sigma}(t)] - \sum_{\alpha}^{N_\alpha} \sum_{k=1}^{N_k} (\boldsymbol{\varphi}_{\alpha k}(t) - \boldsymbol{\varphi}_{\alpha k}^\dagger(t)) \tag{11}$$

$$i\dot{\boldsymbol{\varphi}}_{\alpha k}(t) = [\mathbf{h}_D(t) - i\gamma_{\alpha k}^+ - \boldsymbol{\Delta}_\alpha(t)]\boldsymbol{\varphi}_{\alpha k}(t) - i[\boldsymbol{\sigma}(t)\mathbf{A}_{\alpha k}^{>+} + \bar{\boldsymbol{\sigma}}(t)\mathbf{A}_{\alpha k}^{<+}] + \sum_{\alpha'}^{N_\alpha} \sum_{k'=1}^{N_k} \boldsymbol{\varphi}_{\alpha k,\alpha'k'}(t) \tag{12}$$

$$i\dot{\boldsymbol{\varphi}}_{\alpha k,\alpha'k'}(t) = -[i\gamma_{\alpha k}^+ + \boldsymbol{\Delta}_\alpha(t) + i\gamma_{\alpha'k'}^- - \boldsymbol{\Delta}_{\alpha'}(t)] \cdot \boldsymbol{\varphi}_{\alpha k,\alpha'k'}(t)$$
$$+ i(\mathbf{A}_{\alpha'k'}^{>-} - \mathbf{A}_{\alpha'k'}^{<-})\boldsymbol{\varphi}_{\alpha k}(t) - i\boldsymbol{\varphi}_{\alpha'k'}^\dagger(t)(\mathbf{A}_{\alpha k}^{>+} - \mathbf{A}_{\alpha k}^{<+}). \tag{13}$$

where $\bar{\boldsymbol{\sigma}}(t) = \mathbf{I} - \boldsymbol{\sigma}(t)$ and $N_k = N_d + N_p$ is the total number of the Padé and Lorentzian poles in the contour integral. We term Eqs.(11-13) and their solutions as the Lorentzian-Padé decomposition scheme.

In Reference [17] we used HEOM to calculate the currents of a one-level system, which are the same as those in J. Maciejko, J. Wang and H. Guo's paper [20]. This calibration ensures a good validity and accuracy of our HEOM method.

### B. Initial state calculation

In the reference [17], we set the time derivatives of all the density (and auxiliary density) matrices to be zero and solve the matrix equations for the initial values or the static solutions. In case of the large number of device orbitals (or the expansion terms in the lead spectra), it is very time-consumable to solve the huge matrix equations, even with the sparse matrix technique. In this paper we develop a new calculation method for the initial state calculation.

First, we calculate density matrix $\boldsymbol{\sigma}_D$ from the formula below

$$\boldsymbol{\sigma}_D = \frac{-1}{\pi} \int_{-\infty}^{+\infty} f_P(E) \text{Im}[\mathbf{G}_D^r(E)] \cdot dE, \tag{14}$$

where $f_P(E)$ is the Padé expansion of Fermi-Dirac function,

$$f_P(E) = \frac{1}{2} + \frac{1}{\beta} \sum_{p=1}^{N_p} (\frac{R_p}{E - \mu - z_p^+/\beta} + \frac{R_p}{E - \mu - z_p^-/\beta}) \,. \tag{15}$$

The Green's function is calculated by

$$\mathbf{G}_D^r(E) = [E \cdot \mathbf{I} + \mathbf{h}_D - \mathbf{\Sigma}_{Lrz}^r(E)]^{-1}, \tag{16}$$

where $\mathbf{\Sigma}_{Lrz}^r(E)$ is the self-energy matrix fitted by the multi-Lorentzian expansion (see details in Sec. II C),

$$\mathbf{\Sigma}_{Lrz}^r(E) = \mathbf{\Sigma}_{Lrz}^R(E) + i \cdot \mathbf{\Sigma}_{Lrz}^I(E) = \sum_{d=1}^{N_d} [\frac{-\mathbf{A}_d/W_d}{(E-\Omega_d)^2 + W_d} + i \cdot \frac{\mathbf{A}_d}{(E-\Omega_d)^2 + W_d}]. \tag{17}$$

As $\mathbf{G}_D^r(E)$ has many singularities near the real axis, which makes $\mathbf{G}_D^r(E)$ have a lot of narrow peaks on the real axis, the accurate calculation for the integral of Eq. (14) needs very fine integration grid. However, if the integrand in Eq. (14) has analytic continuation into the upper complex plane, $\mathbf{G}_D^r(E)$ can behave very smooth [21]. So we can do the integral on the upper complex plane with the help of residue theorem [22]. To construct an analytic integrand, we rewrite Eq. (14) as

$$\boldsymbol{\sigma}_D = \frac{-1}{\pi} \text{Im}\{[\int_{-\infty}^{+\infty} f_P(E) \cdot \mathbf{G}_D^r(E) \cdot dE]\} = \frac{-1}{\pi} \text{Im}[\mathbf{I}_\sigma] \tag{18}$$

where $\mathbf{I}_\sigma$ is analytic and it can be calculated by the contour integral as,

$$\mathbf{I}_\sigma = \int_{-\infty}^{+\infty} f_P(E) \mathbf{G}_D^r(E) \cdot dE = -\int_{C_R} f_P(E) \mathbf{G}_D^r(E) \cdot dE + [2\pi i \cdot \sum_k \text{Residue}(k)], \tag{19}$$

where $C_R = C_{R1} + C_{R2} + C_{R3}$, are the integral paths in the upper complex plane, as shown in Fig.1. Residue($k$) is the $k^{\text{th}}$ pole in the contour. As all the poles of $\mathbf{G}^r(E)$ are in the lower complex plane [21], only the Padé poles of $f_P(E)$ are accounted here. After calculating out $\mathbf{I}_\sigma$, $\boldsymbol{\sigma}_D$ is obtained from Eq. (18).

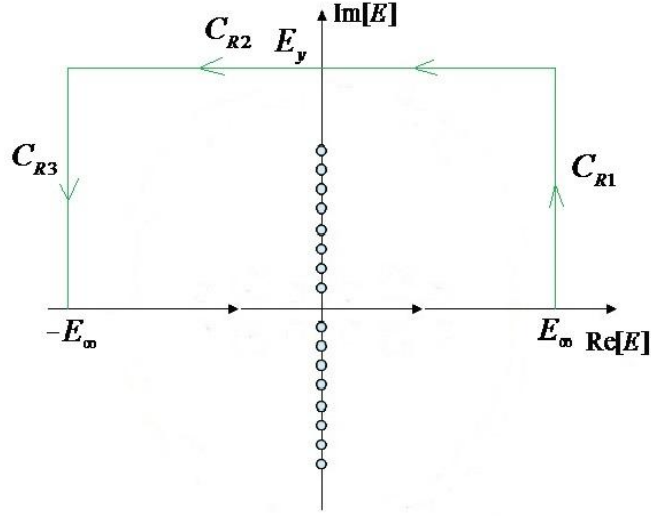

Fig. 1 The contour path and poles of the contour integral in Eq.(19). The filled dots on the y axis represent the Padé poles.

Second, we calculate the 1$^{st}$ tier ADM from its definition (Eq.(9b)). In the equilibrium state ($\Delta_\alpha(t) = 0$), we have

$$\mathbf{G}_D^x(t-\tau) = \frac{-s_x}{\pi} \int_{-\infty}^{+\infty} f_x(E) \operatorname{Im}[\mathbf{G}_D^r(E)] dE$$

$$\mathbf{\Sigma}_{\alpha k}^x(\tau - t) = \mathbf{A}_{\alpha k}^{x,+} \cdot e^{-\gamma_{\alpha k}^+(t-\tau)},$$

where $x = <$ or $>$, $s_< = 1$, $s_> = -1$, $f_<(E) = f_P(E)$, $f_>(E) = 1 - f_P(E)$. Substituting them into $\boldsymbol{\varphi}_{\alpha k}(t)$ definition of equilibrium state as the following (in which case all two-time quantities are reduced to one-time quantities)

$$\boldsymbol{\varphi}_{\alpha k}(t) = i \int_{-\infty}^{t} d\tau [\mathbf{G}_D^<(t-\tau) \cdot \mathbf{\Sigma}_{\alpha k}^>(\tau - t) - \mathbf{G}_D^>(t-\tau) \cdot \mathbf{\Sigma}_{\alpha k}^<(\tau - t)],$$

and doing the integration, we achieve the following result after some derivations

$$\boldsymbol{\varphi}_{\alpha k}(t=0) = \frac{1}{\pi} \int_{-\infty}^{+\infty} dE \frac{\operatorname{Im}[\mathbf{G}_D^r(E)] \cdot [\mathbf{A}_{\alpha k}^{>+} f_P(E) + \mathbf{A}_{\alpha k}^{<+}(1 - f_P(E))]}{\gamma_{\alpha k}^+ + iE}. \tag{20}$$

Similarly contour integral is employed to this integration, for a high accuracy and small computation load. Appendix A gives the details of this.

Third, after solving out $\boldsymbol{\sigma}_D$ and $\boldsymbol{\varphi}_{\alpha k}$, the 2$^{nd}$ tier ADM ($\boldsymbol{\varphi}_{\alpha k, \alpha' k'}$) at equilibrium state is obtained directly from Eq.(13),

$$\boldsymbol{\varphi}_{\alpha k, \alpha' k'} = \frac{1}{\gamma_{\alpha k}^+ + \gamma_{\alpha k}^-} [(\mathbf{A}_{\alpha' k'}^{>-} - \mathbf{A}_{\alpha' k'}^{<-}) \boldsymbol{\varphi}_{\alpha k} - i \boldsymbol{\varphi}_{\alpha' k'}^\dagger (\mathbf{A}_{\alpha k}^{>+} - \mathbf{A}_{\alpha k}^{<+})]. \tag{21}$$

## C. Matrix-based Lorentzian expansion

In the reference [17] we do the Lorentzian fitting for the surface Green's function of the leads. However, we find that for the real or large model system, the number of non-zero self-energy functions is much less than that of the surface Green's functions. So in this paper we directly fit the imaginary part of self-energy matrix functions.

$$\text{Im}[\Sigma_\alpha^r(E)]_{i,j} \approx \sum_{d=1}^{N_{i,j}} \frac{\eta_d}{(E-\Omega_d)^2 + W_d^2}.$$

After fitting all the matrix elements, we use a global index to combine all these Lorentzians together and put the amplitude $\eta_d$ into an amplitude matrix $\mathbf{A}^{(d)}$,

$$\text{Im}[\mathbf{\Sigma}_\alpha^r(E)] = \sum_{d=1}^{N_d} \frac{1}{(E-\Omega_d)^2 + W_d^2} \cdot \mathbf{A}^{(d)}. \qquad (22)$$

The Kramers-Kronig relation is used to get the real part of self-energy [17]. This is the approximation form of the self-energy matrix in Eq. (17). At last, the linewidth function is related to the self-energy by $\mathbf{\Lambda}_\alpha(E) = -2\,\text{Im}[\mathbf{\Sigma}_\alpha^r(E)]$.

## D. Automatic Lorentzian fitting

We use a least-square (LS) solver MINPACK in the Lorentzian fitting. MINPAKC is a software package for solving the nonlinear equations and non-linear LS problems by the Powell's hybrid algorithm and Levenberg-Marquardt algorithm respectively [23]. For the LS problem, we use the subroutine 'lmdif1' in MINPACK to find all the values of $x_j$ (j=1,$n$) for $m$ 'deviation-functions' $f_i$ (i=1,m, m>n) by minimizing the following formula

$$\min\{\sum_{i=1}^{m} f_i(x)^2 : x \in R^n\}.$$

Iteration method is employed to find the optimized parameters from some initial guess.

In our fitting problem, a series of discrete values of self-energy ($S_i = \text{Im}[\Sigma_\alpha^r(E_i)]$) are calculated as the fitting object with the principle layer method [24]. Then the deviation function is obtained after setting the initial fitting parameters $\{x_j\}$,

$$f_i(x_1, x_2, \cdots x_n) = S_i - L(E_i, x_1, x_2, \cdots x_n),$$

where $L$ is the multi-Lorentzian function,

$$L(E, x_1, x_2, \cdots x_n) = L(E, \Omega_1, W_1, \eta_1, \Omega_2, W_2, \eta_2, \cdots \eta_{N_d}) = \sum_{d}^{N_d} \frac{\eta_d}{(E-\Omega_d)^2 + W_d^2}.$$

$\{x_j\}$ correspond to the Lorentzian parameters: $\{\Omega_d, W_d, \eta_d\}$, so $n = 3N_d$.

Since $n$ is very large, $\sum_{i=1}^{m} f_i(x)^2$ has a lot of local minimums in such high-dimensional solution space ($\{x_j, x \in R^n\}$). Different initial guess leads different final solutions. It is very important to set proper initial Lorentzian parameters for a good fitting result. Some codes are wriiten to automatically find the initial Lorentzian parameters.

After fitting all the $N_z$ curves, we have

$$\text{Im}[\Sigma^{(1)}(E)] = \sum_{d=1}^{N_d^1} \frac{\eta_d^{(1)}}{(E - \Omega_d^{(1)})^2 + W_d^{(1)\,2}} \;;\; \text{Im}[\Sigma^{(2)}(E)] = \sum_{d=1}^{N_d^2} \frac{\eta_d^{(2)}}{(E - \Omega_d^{(2)})^2 + W_d^{(2)\,2}} \;;\ldots\ldots$$

where $N_d^k$ is the number of Lorentzians for the $k^{th}$ curve. For small systems, we directly combine all these Lorentzians into Eq.(22) by defining a global Lorentzian index. The total number of Lorentzians is $N_{d0} = \sum_{k=1}^{N_z} N_d^k$. We term this as *scheme-0* fitting method.

For large systems, the number of Lorentzians has to be reduced. Following part gives a detailed description.

Since the shape of a Lorentzian is mainly determined by its center ($\Omega$) and width ($W$), we present each Lorentzian function as one point on a $W$-$\Omega$ diagram. If some points are closely located, we can combine them into one new point (see Appendix B). After this combination, the number of Lorentzians is reduced from $N_{d0}$ to $N_{d1}$. Then we refit all $\text{Im}[\Sigma_\alpha^r(E)]_k$ curves again for optimization.

One refitting way is to maintain all the combined $\{W_d\}$ and $\{\Omega_d\}$ values and only to fit the amplitude $\{\eta_{k,d}\}$, where $k = 1, \cdots N_z$, is the index for different curves and $d = 1, \cdots N_{d1}$ is the number of combined Lorentzians. The number of fitting parameters is $N_z N_{d1}$. It is noted that for each curve-fitting, $\{\Omega_d\}$ and $\{W_d\}$ come from the combined Lorentzian points, which is much more than those in the original curve fitting. So the fitting results are often much better. We term this process as *scheme-1* fitting method.

Another refitting way is to treat the combined $\{W_d\}$ and $\{\Omega_d\}$ as the initial fitting parameters. Their values are optimized together with the amplitudes in the refitting. We combine all the curves into one large fitting object ( $S = \{\text{Im}[\Sigma_1(E_i)], \text{Im}[\Sigma_2(E_i)], \cdots \text{Im}[\Sigma_{N_z}(E_i)]\}$ ) and simultaneously fit all these curves with one set of Lorentzian parameters $\{\Omega_d, W_d, \eta_{k,d}\}$. In other words, we minimize the following quantity by the LS algorithm:

$$SRR = \sum_{i=1}^{N} | S_1(i) - \sum_{d=1}^{N_d} \frac{\eta_{1,d}}{(E_i - \Omega_d)^2 + W_d^2} |^2 + \sum_{i=1}^{N} | S_2(i) - \sum_{d=1}^{N_d} \frac{\eta_{2,d}}{(E_i - \Omega_d)^2 + W_d^2} |^2 + \cdots\cdots$$
$$+ \sum_{i=1}^{N} | S_{N_z}(i) - \sum_{d=1}^{N_d} \frac{\eta_{N_z,d}}{(E_i - \Omega_d)^2 + W_d^2} |^2$$

In this refitting process not only the $N_z N_{d1}$ amplitudes $\{\eta_{k,d}\}$, but also the center $\{\Omega_d\}$ and width $\{W_d\}$ are to be adjusted. The number of fitting parameters is $2N_{d1} + N_z N_{d1}$. We term this type of global curve fitting as the *scheme-2* fitting method.

Among these three fitting schemes, scheme-2 can produce the fewest Lorentzians and the fitting results are best in most cases. But since the fitting object is very large ($N$ curves with $N \cdot N_z$ data points or object functions), the fitting time is usually much longer than the other two schemes.

### III. Numerical Implementation for GNR

Graphene nanoribbons are used as HEOM application examples in this paper. The nearest neighbor tight-binding model is employed with one $p_z$ orbital for each carbon atom. The GNRs are classified into armchair type and zigzag type (denoted as AGNR and ZGNR). So we treat these two types GNR respectively.

#### A. Armchair-type GNR calculation

In this section, AGNR (M=16) is used. M means the number of atoms in one GNR unit cell. (In other graphene papers, this GNR is often noted as 8-AGNR).

Fig.2(a) shows the atom configuration and the device atom labels of this AGNR. In the nearest TB model, there are 20 non-zero self-energy curves. To reduce the number of Lorentzians, we find out all the different curves because there are some degenerate curves due to the geometry symmetry of this GNR. Taking out the degeneracy, 10 different self-energy curves are left. Fig.2(b) shows 4 of them (the imaginary part). We see that these curves are of complicated shapes. A lot of Lorentzians are needed to fit them. The fitting results determine the accuracy of HEOM calculation. Two schemes are used to fit the self-energy curves of this AGNR.

(A1) The delta peak in the self-energy curve: a type of van-Hove singularity

Before fitting, we notice that near the energy range of E=0 eV, the real parts of the self-energies are very large (which can be infinite) while the imaginary parts are of finite values (as shown in Fig.2 (c)). This behavior differs from the fact that real and imaginary parts should obey the Kramers-Kronig relation [25].

To find out the reason, we use a fine energy sampling which is comparable to the small imaginary number ($i\eta$) in the surface green's function calculation [4]. We find in each imaginary-part curve there exist a 'delta-function-like' peak in the range of E=0 (Inset of Fig. 2(c)). This is reasonable because by the Kramers-Kronig relation, $\text{Re}[\Sigma(E)]$ can be transformed from $\text{Im}[\Sigma(E)]$:

$$\text{Im}[\Sigma(E)] = -A_{xx}\pi \cdot \delta(E-E_0), \quad \text{Re}[\Sigma(E)] = \frac{1}{\pi}\int \frac{\text{Im}[\Sigma(E)]}{E'-E}dE' = \frac{A_{xx}}{E-E_0}.$$

So the $\text{Re}[\Sigma(E)]$ curves have the fractional-function-like peak, as shown in Fig.2 (c). This explains the large behavior of the real-part curves at E=0.

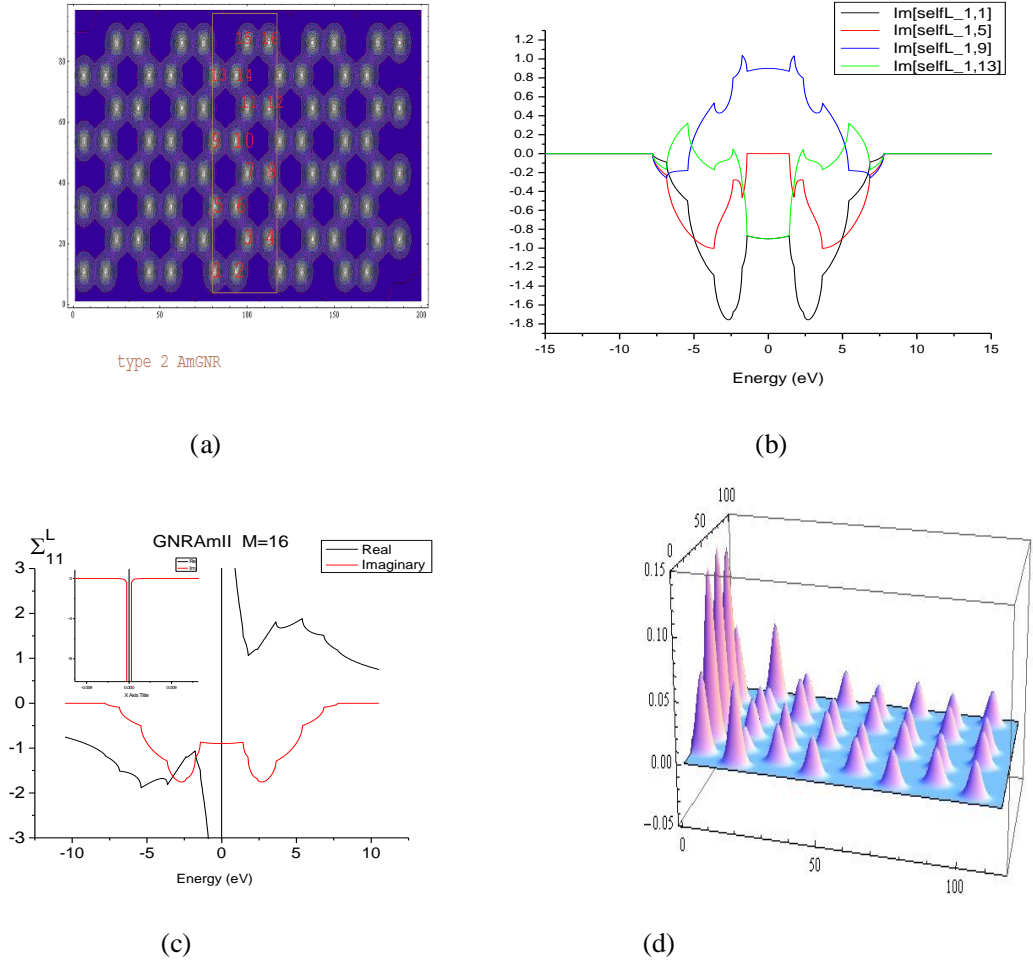

Fig.2 (a) Atomic configuration and atom index for the AGNR (M=16). The rectangular region denotes the device. (b) 4 examples of self-energy curves (imaginary part) of this AGNR. (c) The imaginary and real part of self-energy ($\Sigma_{1,1}$). The inset show the magnified delta-function-like peak near the E=0 region. (d) The local density of states distribution for the edge-state in the AGNR lead.

We know in the 'edge-state' of zigzag GNR, there is a flat band in the band-structure at E=0, whose DOS curve has a sharp peak [26-27]. This is very similar to the delta-peak discussed here. In fact, on the terminal side of such semi-infinite armchair GNR, the atom configuration is also zigzag (see Fig.2 (a)). So we guess such delta peak state in semi-infinite AGNR is another form of edge state.

To demonstrate this, we plot the local density of states (LDOS) at E=0 for this semi-infinite AGNR, as shown in Fig. 2(d). We see this state decays exponentially from the terminal to the inside. (The LDOS does not decay to zero, but a constant, which is the normal state near E=0 region). From our knowledge, this edge state in the semi-finite AGNR is reported for the first time.

In the fitting, a narrow Lorentzian (i.e. $\frac{A}{(E-E_0)^2+W^2}$, the width W is a very small value, like 0.00001 eV) is used to stand for such delta peak. To evaluate the amplitude of this narrow Lorentzian, following steps are used: (1) Obtain $A_{xx}$ in from the real-part curve. (2) Assume that the integrals for both the delta function and the narrow Lorentzian are equal,

$$\frac{-A\pi}{W} = \int_{-\infty}^{+\infty} \frac{-A}{(E-\Omega)^2+W^2} dE = \int_{-\infty}^{+\infty} -A_{xx}\pi \cdot \delta(E) dE = -A_{xx}\pi.$$

Thus the Lorentzian amplitude is obtained: $A = A_{xx}W$.

(A2) Scheme-1 Lorentzian fitting and the HEOM calculation

We use the scheme-1 method to fit all the self-energy curves of this AGNR. As stated before, 10 different curves are fitted. The parameter $M_{peak}=2$ is used, which means 2 Lorentzians are used for one peak of every curve. The fitting results are listed in Table 1 below. For each curve, two sets of initial Lorentzian parameters are used in optimization. In one initial parameter set, the Lorentzians are uniformly positioned in each peak and in another initial parameter set they are positioned non-uniformly. These two initial conditions result in two optimization results. The better one is chosen.

| Curve index | SSR1 | SSR2 | SSR3 | SSR4 | SSR5 | SSR6 | $N_{Dpeak}$ |
|---|---|---|---|---|---|---|---|
| 1 | 425.006 | 3.250 | 425.006 | 1.961 | 425.006 | 0.111 | 1 |
| 2 | 138.618 | 1.995 | 138.618 | 0.846 | 138.618 | 0.180 | 1 |
| 3 | 122.511 | 0.148 | 122.511 | 0.286 | 122.511 | 0.092 | 1 |
| 4 | 64.994 | 0.113 | 64.994 | 0.088 | 64.994 | 0.126 | 1 |
| 5 | 592.178 | 4.848 | 592.178 | 1.213 | 592.178 | 0.287 | 1 |
| 6 | 94.575 | 1.166 | 94.575 | 0.259 | 94.575 | 0.185 | 1 |
| 7 | 45.549 | 0.245 | 45.549 | 0.183 | 45.549 | 0.290 | 1 |
| 8 | 670.647 | 5.732 | 670.647 | 3.420 | 670.647 | 0.232 | 1 |
| 9 | 166.712 | 1.214 | 166.712 | 1.154 | 166.712 | 0.132 | 1 |
| 10 | 938.583 | 12.744 | 938.583 | 3.191 | 938.583 | 0.296 | 1 |

Table 1 The Lorentzian fitting results (column 2-7) and the number of delta-peak ($N_{Dpeak}$) (column 8) for each different self-energy curve of AGNR (M=8). The results are measured by the sum of squared residues (SSR)..
SSR1: SSR before optimization by the 1st type initial Lorentzian parameters;
SSR2: SSR after optimization by the 1st type initial Lorentzian parameters;
SSR3: SSR before optimization by the 2nd type initial Lorentzian parameters;
SSR4: SSR after optimization by the 2nd type initial Lorentzian parameters;
SSR5: SSR before optimization with the combined Lorentzian points;
SSR6: SSR after optimization with the combined Lorentzian points..

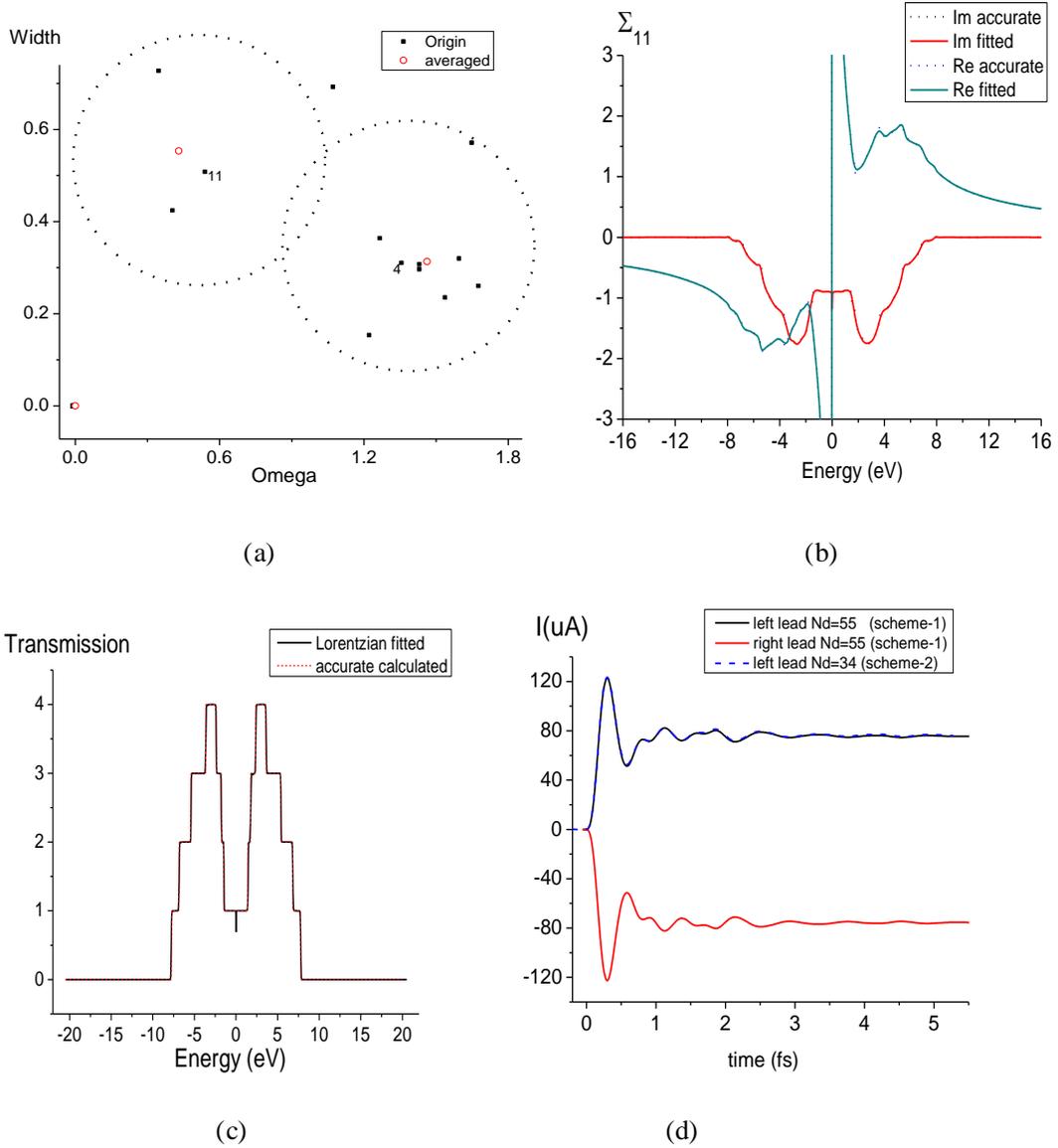

(a)                                       (b)

(c)                                       (d)

Fig.3 (a) Combination process of the Lorentzian points in Omega-Width diagram. The original fitted Lorentzians are presented as the black square dots and the red round dots are the average combined points.. The two dotted circles mean the combination region for each center points (labeled as 4 and 11). $N_d$ is reduced from 130 to 55 after the combination. (b) The fitted and accurate self-energy curve (real and imaginary part) of $\Sigma_{1,1}$. (c) The transmission spectrum of AGNR(M=16) from the accurate calculation (dotted line) and the schem-1 Lorentzian fitting method (solid curve). (d) The transient currents of AGNR(M=16) (left current: black line; right current: red line). The dotted line is for the left current calculated from the scheme-2 fitting method, as discussed in Sec. III A3.

After fitting all the curves, totally 130 Lorentzians are obtained. Then we plot these Lorentzians on the $W$-$\Omega$ diagram and combine the points for reducing the number of Lorentzians. Fig. 3(a) shows one magnified part of this diagram for the combination process. In Fig. 3(a), black dots represent the originally fitted Lorentzian points and the red dots represent the combined points. An algorithm is used to effectively group all the original neighboring points into the combined points, which is detailed in Appendix B. With the combination parameter rr=0.4, the number of Lorentzian points are reduced to 55. In scheme-1 method, we fix these 55 pairs of $\{\Omega_d\}$ and $\{W_d\}$ as the new fitting parameters and refit all the curves by adjusting the amplitudes $\{\eta_{k,d}\}$. Fig. 3(b) shows one example of the results. We see both the imaginary and the real parts of $\Sigma_{1,1}$ are fitted well. In Table 1, column 6(7) lists the LS results before (after) the amplitude refitting. We see the refitting results are much better than the initial fitting. It is reasonable since the number of Lorentzians in the refitting process is larger than the original one.

With the fitting parameters: $\{\Omega_d\}$ and $\{W_d\}$ and $\{\eta_{k,d}\}$, all the 10 self-energy curves are recovered. Through the NEGF formula [4], the transmission spectrum is obtained, which agrees well with the accurate curve, as shown in Fig.3 (c).

With the method in Sec.II B for the initial state, the dynamic currents of this system are calculated, as shown in Fig. 3(d). A bias voltage of 2 V is applied across the device region and the on-site energy linear changes between the two leads. The bias voltage varies as a step function at t=0 in time domain. We see after some oscillations in the early time, the currents tend to a steady value at about 4 fs.

(A3) Scheme-2 Lorentzian fitting and the HEOM calculation

Similarly, we also use the global fitting method—scheme-2 to this AGNR as a comparison. Fig. 4(a) shows the Lorentzian points in the $W$-$\Omega$ diagram in different steps of such fitting scheme. The left plot shows the Lorentzian points from the original fitting ($N_d$=108, without the delta-peak point). The middle plot shows the combined Lorentzian points ($N_d$=33, without the delta-peak point). The right plot shows the globally fitted Lorentzian points ($N_d$=34, with the delta-peak point). With this set of Lorentzian parameters, we calculate the transmission spectrum, which is shown by the black-solid curve in Fig. 4(b). It is very close to the accurate transmission curve (the red-dotted curve in Fig. 4(b)). So the fitting results are very good in the scheme-2 method. The computation time for this global fitting is about 10 minutes (in one 2.6 GHz CPU).

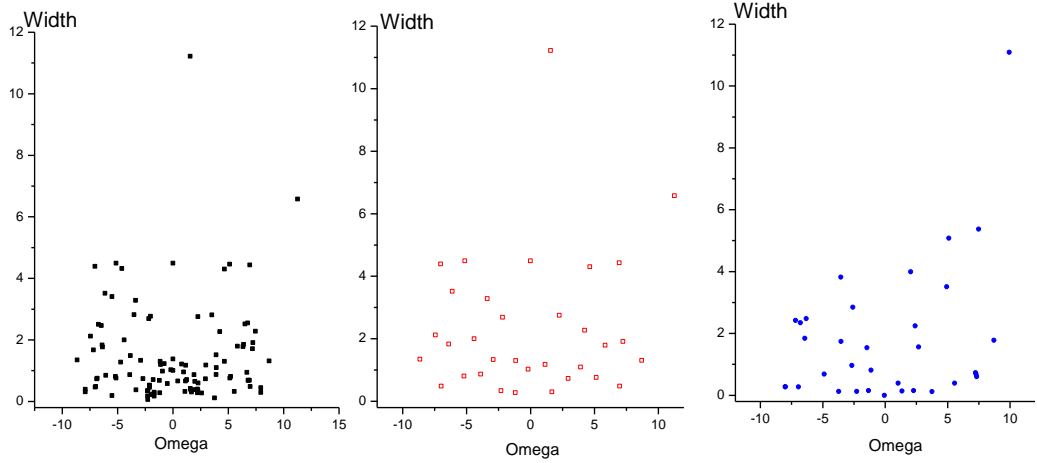

(a)

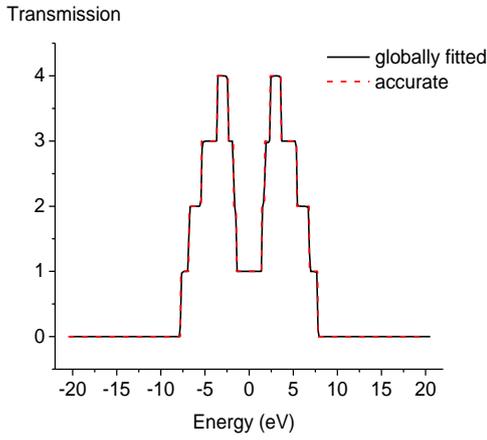

(b)

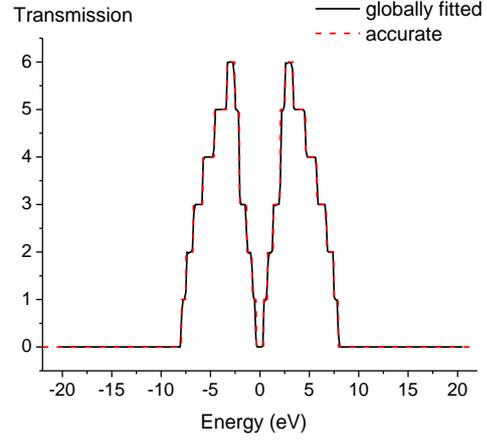

(c)

Fig.4 (a) Lorentzian points in different steps of Scheme-2 fitting for AGNR (M=16). The left plot shows the points after the original fitting; the middle plot shows the combined points; the right plot shows the points after the global fitting process. (b) The transmission spectrum of the AGNR(M=16) by the accurate calculation (dotted line) and the scheme-2 Lorentzian fitting method (black line). (c) The transmission spectrum of the AGNR(M=24) by the accurate calculation (dotted line) and the scheme-2 Lorentzian fitting method (black line).

We also use scheme-2 method to fit another larger AGNR with M=24 atoms in each unit cell. The number of Lorentzian points is reduced from 384 to 41 after the point combination. The transmission spectrum calculated by this fitting scheme agrees well with spectrum by the accurate calculation, as shown in Fig.4(c).

## B. Zigzag-type GNR calculation

Now we turn to the zigzag-type GNR. For ZGNR, there are several new properties in the Lorentzian fitting and HEOM calculations.

(B1) Fitting for the narrow peaks

We choose M=8 ZGNR as an example (Fig. 5(a) shows the atom configuration and labels in the device region). 12 different self-energy curves are obtained from 20 non-zero self-energy elements (Fig. 5(c) shows first 5 of them). In each of the self-energy curves there exist several (2-5) very narrow peaks. These narrow peaks can only be fitted by very narrow Lorentzians. From the Kramers-Kronig relation, we know a very narrow Lorentzian peak in the imaginary part of a function will lead to a very large Lorentzian peak in the real part. This can be seen below:

$$\text{Im}[\Sigma^r] \approx \sum_d \frac{A_d}{(E-\Omega_d)^2 + W_d^2} \qquad \text{Re}[\Sigma^r] \approx \sum_d \frac{-A_d/W_d \cdot (E-\Omega_d)}{(E-\Omega_d)^2 + W_d^2}.$$

We see a small $W_d$ leads to a large $\text{Re}[\Sigma^r]$.

Thus a tiny fitting error in the imaginary part will result in a very huge error in the real part. So in the fitting, very narrow Lorentzians (W<1.0d-4) must be avoided to ensure good behaviors of the corresponding real-part curves.

If some very narrow Lorentzians are found in fitting, we have to change the initial fitting conditions and refit the curve, until all the Lorentzian widths are large enough. This increases the computation time. Further more, for some complicated curves it is very difficult to find the proper wide-width Lorentzians.

Here a new strategy is employed to effectively fit these narrow-peak curves. Since very narrow Lorentzians must be used, we assign some fixed values for their small widths. These widths will no long be involved in the fitting. For example, for a curve with 3 narrow peaks, we use 6 (in each peak 2 Lorentzians are used) narrow Lorentzians with a fixed width of 0.01 eV to fit the narrow peaks. Only their centers and amplitudes ($\{A_d, \Omega_d\}$ ($d=1,\cdots 6$)) will be fitted. For other normal peaks, their Lorentzian centers, widths and amplitudes are all the fitting parameters. This is shown below.

$A_1, \Omega_1, W_s(0.01)$
$A_2, \Omega_2, W_s(0.01)$
$\vdots$
$A_6, \Omega_6, W_s(0.01)$
$A_7, \Omega_7, W_7$
$\vdots$
$A_N, \Omega_N, W_N$

The fitting parameters are: { $A_1, \Omega_1, A_2, \Omega_2, \cdots, A_6, \Omega_6$ ; $A_7, \Omega_7, W_7, \cdots, A_N, \Omega_N, W_N$ }.

(B2) Scheme-2 Lorentzian fitting and HEOM calculations

We use scheme-2 method to fit M=8 ZGNR. The number of Lorentzian points in the original fitting process is 373 ($M_{peak}$=3). After combination the number of Lorentzian points is reduced to 127 (rr=0.2). Then the global fitting is done for all the 12 different self-energy curves(imaginary part). Due to the narrow peaks, $N_d$ is much larger and the fitting time here is much longer (about 2.7 hours in a 2.6 GHz CPU) compared to that of AGNR (Sec. III A3). Fig, 5(c) shows first 5 self-energy curves as an example (with the shifted energy [28]). We see both for the imaginary part and real part, the curves from Lorentzian fitting and accurate calculations agree well. With these fitting parameters, the transmission spectrum also agrees well with the accurate one, as shown in Fig. 5(d). With the static state HEOM calculation, the dynamic current induced by a step-function bias voltage (1.0 V) is calculated, as shown in Fig. 5(e). We see a very small steady current is achieved for this ZGNR.

We also use another combination criteria (rr=0.1) in the scheme-2 fitting for this ZGNR. In this new fitting, the number of combined Lorentzian points ($N_d$) is 166. With the same HEOM approach, the dynamic currents are obtained and shown as the dotted lines in Fig. 5(e). We see the two fittings produce very similar current curves, which ensures the convergence of our Lorentzian fitting scheme.

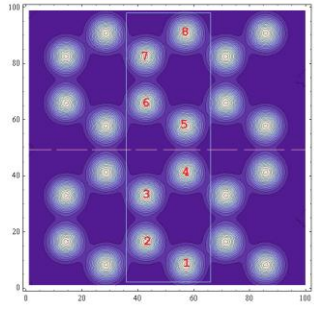

(a)

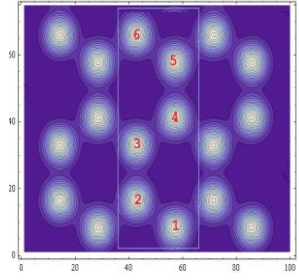

(b)

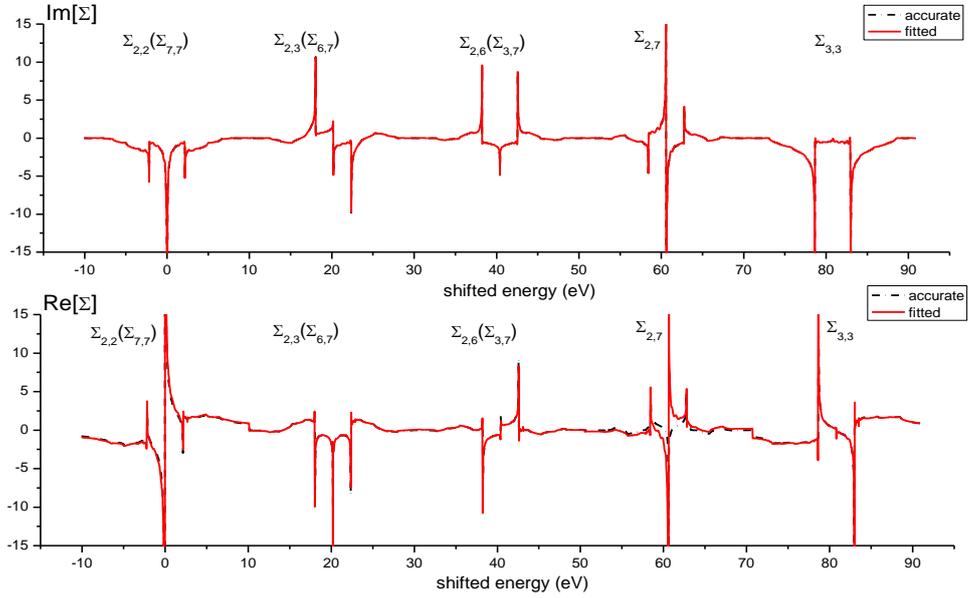

(c)

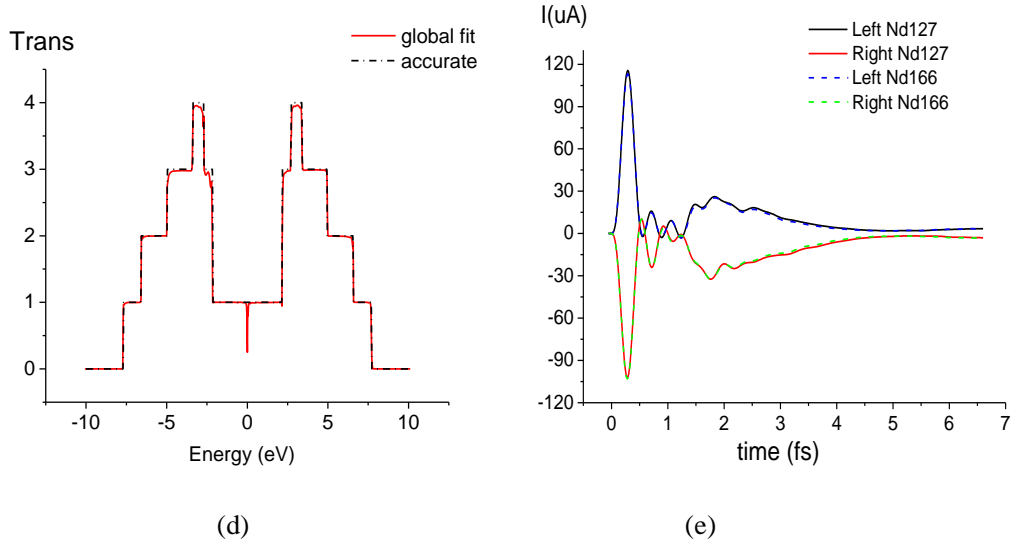

(d)                          (e)

Fig. 5 (a) Atom configuration of ZGNR (M=8). The rectangular box denotes the device region. The dotted line means the mirror symmetric line for the even-odd effect (see Sec. III B3). (b) Atom configuration of ZGNR (M=6). The rectangular box denotes the device region. (c) The first 5 different self-energy curves of ZGNR (M=8) from the accurate calculation (dotted line) and scheme-2 Lorentzian fitting method (solid line). The upper panel is for the imaginary part and the lower panel is for the real part. (d) The transmission spectrum of ZGNR (M=8) from the scheme-2 Lorentzian fitting (solid line) and the accurate calculation (dotted line). (e) The Transient current of ZGNR (M=8) by two scheme-2 Lorentzian fittings. The on-site energy in the device is linearly changed between two leads under a bias of 1.0 V. The parameter $N_d$ is 127 (Solid lines) or 166 (dotted lines).

(B3) Dynamic currents of ZGNR with even-odd effect

    When comparing the transmission spectrum and the dynamic current of this ZGNR (Fig. 5(d) and Fig.5 (e)), we find that since the transmission is 1.0 near the zero energy, the steady value of current approaches to zero at a bias of 1.0 V. These two figures seem inconsistence due to the Landauer formula [4]. In fact the transmission spectrum in Fig. 5(d) is under a zero bias voltage. If a finite bias voltage is applied on this ZGNR, the transmission spectrum will have a sharp deep around E=0 with a width of such bias value (See Fig.6 (a)). The new transmission spectrum is consistent with the current result. This phenomenon is first observed by Z.Y.Li et al and is explained as the parity mismatch of bonding and antibonding orbitals [29]. They called this as the even-odd effect because only the symmetric ZGNR have such effect.

    For the asymmetric ZGNR, such as the M=6 case (see Fig. 5(b)), there is no parity for the orbitals and the transmission curve does not drop to zero under a bias voltage (see Fig. 6(b)). For M=6 ZGNR, we also use our scheme-2 fitting approach for the HEOM calculation. The transient current under a bias of 1.0 V is shown in Fig. 6(c). As expected, the steady current has a finite value, not zero, due to the even-odd effect.

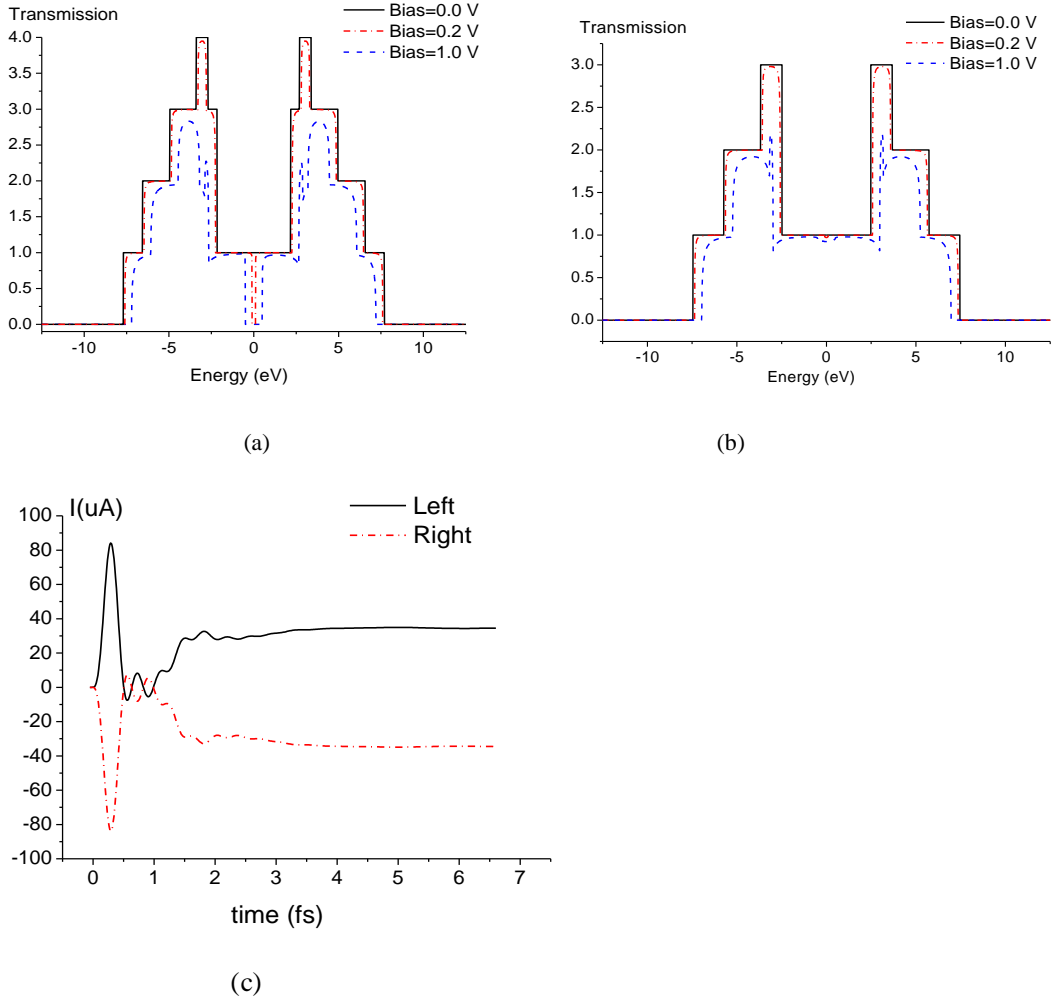

(c)

Fig. 6 (a) The transmission spectrum of ZGNR (M=8) with a bias voltage of 0 V (solid line), 0.2 V (dotted line) and 1.0 V (dashed line). (b) The transmission spectrum of ZGNR (M=6) with a bias voltage of 0 V (solid line), 0.2 V (dotted line) and 1.0 V (dashed line). (c) The transient current of the left (solid line) and right (dashed line) lead in ZGNR (M=6) after a step-wise bias pulse of 1.0V at time=0. The on-site energy in the device is linearly changed between two leads.

## IV. Conclusion

HEOM have been developed in two aspects in this paper. One is the energy integration method for the initial state calculation, including the contour integral for accurate computation. Another is a systematic set of Lorentzian fitting schemes. With these schemes, the number of Lorentzians can be effectively reduced by the Lorentzian combination algorithm; at the same time the fitting accuracy is still maintained by the refitting process or the preestablished narrow widths for the sharp peaks. Armchair and zigzag graphene nanoribbons are used as calculation examples. We find the delta-function-like states in many semi-infinite AGNR, which contribute to the real part of self-energies. Our HEOM calculation also shows the currents evolution in ZGNR with the even-odd effect for the first time.

## Acknowledgement

We would like to thank Dr. ChiYung Yam and Dr. Hui Cao for some useful discussions on the quantum transport and Dr. Jian Sun for the help with the computer. Support from the Hong Kong Research Grant Council (HKU700808P, HKU700909P, HKU700711P, HKUST9/CRF/08), AoE (AOE/P-04/08), National Science Foundation of China (21103157, 21033008), and Fundamental Research Funds for the Central Universities of China (2340000034) is gratefully acknowledged.

# Appendix
## Appendix A. Initial state calculation of the 1st tier ADM by contour integration

Similar to the density matrix calculation in Sec. II B, we use the residue theorem to the integral calculation of the 1st tier ADM. We rewrite Eq.(20) as:

$$\varphi_{\alpha k} = \frac{1}{\pi i} \int_{-\infty}^{+\infty} dE \frac{\text{Im}[\mathbf{G}_D^r(E)] f_P(E) \cdot [\mathbf{A}_{\alpha k}^{>+} - \mathbf{A}_{\alpha k}^{<+}] + \text{Im}[\mathbf{G}_D^r(E)] \cdot \mathbf{A}_{\alpha k}^{<+}}{E - i\gamma_{\alpha k}^+} = \frac{1}{\pi i}(\mathbf{I}_a + \mathbf{I}_b)$$

where $\mathbf{I}_a = \int_{-\infty}^{+\infty} dE \frac{\text{Im}[\mathbf{G}_D^r(E)] f_P(E) \cdot [\mathbf{A}_{\alpha k}^{>+} - \mathbf{A}_{\alpha k}^{<+}]}{E - i\gamma_{\alpha k}^+}$, $\mathbf{I}_b = \int_{-\infty}^{+\infty} dE \frac{\text{Im}[\mathbf{G}_D^r(E)] \cdot \mathbf{A}_{\alpha k}^{<+}}{E - i\gamma_{\alpha k}^+}$.

Since only the analytic function can be used in the residue theorem, we have to change the integrand above (like $\frac{\text{Im}[\mathbf{G}_D^r(E)] \cdot \mathbf{A}_{\alpha k}^{<+}}{E - i\gamma_{\alpha k}^+}$) as Re + iIm form:

$$\frac{1}{E - i\gamma_{\alpha k}^+} = \frac{1}{E - i(\gamma_1 + i\gamma_2)} = \frac{1}{E + \gamma_2 - i\gamma_1} = \frac{(E + \gamma_2) + i\gamma_1}{(E + \gamma_2)^2 + \gamma_1^2},$$

where $\gamma_1$ and $\gamma_2$ is the real and imaginary part of $\gamma_{\alpha k}^+$. So the two integrands above can be written as the form

$$\text{Im}[G_D^r](g_1 + ig_2) = \text{Im}[G_D^r]g_1 + i\text{Im}[G_D^r]g_2 = \text{Im}[G_D^r g_1] + i\text{Im}[G_D^r g_2]$$

where $g_1$ and $g_2$ stand for the real functions in the integrands except $\text{Im}[G_D^r]$.

Then the two integrals above can be rewritten as the form below

$$\int \text{Im}[G_D^r](g_1 + ig_2)dE = \int \text{Im}[G_D^r]g_1 dE + i\int \text{Im}[G_D^r]g_2 dE = \text{Im}[\int G_D^r g_1 dE] + i\text{Im}[\int G_D^r g_2 dE]$$

The integrals in Im[] above are analytic, which can be used for the residue theorem. For example, $\mathbf{I}_a$ can be written as

$$\mathbf{I}_a = \text{Im}[\int_{-\infty}^{+\infty} dE \frac{\mathbf{G}_D^r(E) f_P(E) \cdot (E + \gamma_2) \cdot [\mathbf{A}_{\alpha k}^{>+} - \mathbf{A}_{\alpha k}^{<+}]}{(E + \gamma_2)^2 + \gamma_1^2}] + i\text{Im}[\int_{-\infty}^{+\infty} dE \frac{\mathbf{G}_D^r(E) f_P(E) \cdot \gamma_1 \cdot [\mathbf{A}_{\alpha k}^{>+} - \mathbf{A}_{\alpha k}^{<+}]}{(E + \gamma_2)^2 + \gamma_1^2}]$$

$$= (\text{Im}[\mathbf{I}_{a1}] + \text{Im}[\mathbf{I}_{a2}]) \cdot [\mathbf{A}_{\alpha k}^{>+} - \mathbf{A}_{\alpha k}^{<+}]$$

The two integrations ($\mathbf{I}_{a1}$ and $\mathbf{I}_{a2}$) can be calculated by the contour integral,

$$\mathbf{I}_{a1} = \int_{-\infty}^{+\infty} dE \frac{\mathbf{G}_D^r(E) f_P(E) \cdot (E+\gamma_2)}{(E+\gamma_2)^2 + \gamma_1^2} = \oint dE \frac{\mathbf{G}_D^r(E) f_P(E) \cdot (E+\gamma_2)}{(E+\gamma_2)^2 + \gamma_1^2} = 2\pi i \sum_k^{N_k} \text{Residue}[k]$$

$$\mathbf{I}_{a2} = \int_{-\infty}^{+\infty} dE \frac{\mathbf{G}_D^r(E) f_P(E) \cdot \gamma_1}{(E+\gamma_2)^2 + \gamma_1^2} = \oint dE \frac{\mathbf{G}_D^r(E) f_P(E) \cdot \gamma_1}{(E+\gamma_2)^2 + \gamma_1^2} = 2\pi i \sum_k^{N_k} \text{Residue}[k],$$

where Residue(k) stands for the residues in the integral contours, which include the Padé poles from $f_P(E)$ and the pole from the fractional function $\frac{1}{(E+\gamma_2)^2 + \gamma_1^2}$. It is easy to prove that integral on upper-complex-plane contours (like $C_R = C_{R1} + C_{R2} + C_{R3}$ in Fig. 1) tends to zero when R tends to infinity, so the integral on real axis can be transformed to the contour integral. Another integral $\mathbf{I}_b$ can also be calculated in the same way.

## Appendix B. Algorithm for Lorentzian points combination

In Sec. II D, some algorithm is used to combine the neighboring Lorentzian points into one point on the Omega-Width diagram. Here shows the details of this algorithm. There are several steps for searching and combining all the Lorentzian points (the total number is N).

(1) Search for each point with index i (i=1,…, N) except the grouped points. The selected point is viewed as center point.
(2) For each center point i, search for the left points with index j (j=i+1,…,N), except for the grouped points.
(3) Calculate the distance Dij between point i and j. If Dij is less than the critical radius rr, then group the point j as the combined point with respect to the center point i. (Array party(i,k)=j and array Ifgroup(j)=1 is used for recording).
(4) Do the search ( j and i loop), until all the points are grouped.
(5) Use the recording information to calculate the average value for each group of points.